\documentclass[a4paper,review,10pt]{elsarticle}
\usepackage{graphicx}
\usepackage{fancyhdr}

\journal{Solid-State Electronics}

\begin{document}

\title{
Capacitances in micro-strip detectors: a conformal mapping approach
}
 
\author{Paolo Walter Cattaneo \\
Paolo.Cattaneo@pv.infn.it \\
INFN Pavia, Via Bassi 6, Pavia, I-27100, Italy }

\begin{abstract}
The knowledge of capacitance in semiconductor micro-strip detectors is important for a correct 
design, simulation and understanding of the detectors.\\
Analytical approaches can efficiently complement numerical methods providing quick results in the
design phase.\\
The conformal mapping method has proved to be the most effective analytical approach providing
many realistic models \cite{cap90}-\cite{cat95}.\\ 
In this paper improved analytical results are presented and compared with experimental data.\\ 
The excellent agreement between predictions and measurements confirms the relevance of 
this approach to modeling realistic detectors.
\end{abstract}


\maketitle

\section{The role of capacitances in micro-strip detectors}

\begin{center}
\begin{figure}
\includegraphics[width=1.1\textwidth,angle=0] {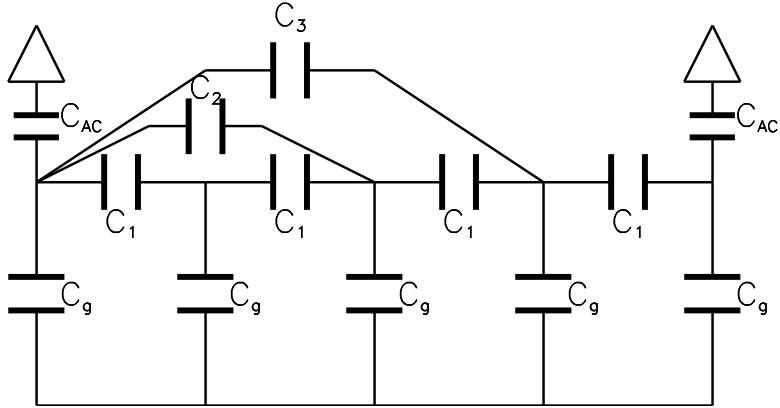}
\caption{Capacitive circuital model of a micro-strip detector section. Every 4th strip is read out, 
employing the principle of capacitive charge division.}
\label{detcap}
\end{figure}
\end{center}

A micro-strip detector can be modeled as a circuit: the energy release is a 
current generator and the resulting signal propagates in a network of capacitances and 
resistances. The nodes of the network are the strips and the backplane.
A realistic model might incorporate also the read out electronics \cite{cat95}.\\
The capacitive elements of the circuital model are shown in Fig.\ref{detcap}: $C_g$ is the 
strip to ground capacitance, while $C_n$ is the interstrip capacitance $n$ strips away.\\ 
The relation between the output signals and the energy release is mediated 
by the detector capacitances.
These are difficult to calculate because of the complex multi-electrode geometry.\\
A numerical solution of the Poisson equation within the detector is possible but is 
of limited use for the design because it has to be repeated for every configuration. Therefore
it is interesting to investigate approximate analytical solutions.

\section{Capacitance calculations with conformal mapping}

\subsection{Conformal mapping}

A powerful approach to calculate electrostatic fields and capacitances in complex geometry with 2D
symmetry is the use of conformal mappings \cite{binns}-\cite{schinziger}-\cite{cap90}.\\
They are angle preserving 1 to 1 functions between regions of the complex plane. This property
is equivalent to the requirement that they are differential as complex functions \cite{alfhors}.\\
An electrostatic field satisfying the Laplace equation in a region with 2D symmetry can 
be described by a complex potential function $F(z=x+i y) = t(x,y) = u(x,y) + iv(x,y)$, where
$u(x,y)$ is the potential function and $v(x,y)$ the flux function defined such that the curves 
with constant $v(x,y)$ and with constant $u(x,y)$ are orthogonal.
Both $v(x,y)$ and $u(x,y)$ satisfy the Laplace equation and $F$ is differential and conformal.\\
Any conformal function preserves potential and flux functions
and the capacitances between corresponding conductors.\\
This property is the key for using conformal mapping in capacitance calculation. 
A region with a non-trivial distribution of conductors (and/or boundary conditions) is
mapped onto a region (possibly through intermediate configurations and exploiting symmetry) 
where the capacitance is known, e.g. parallel plates.

\subsection{Schwartz-Christoffel transformations}

\begin{center}
\begin{figure}
\includegraphics[width=9cm,angle=0] {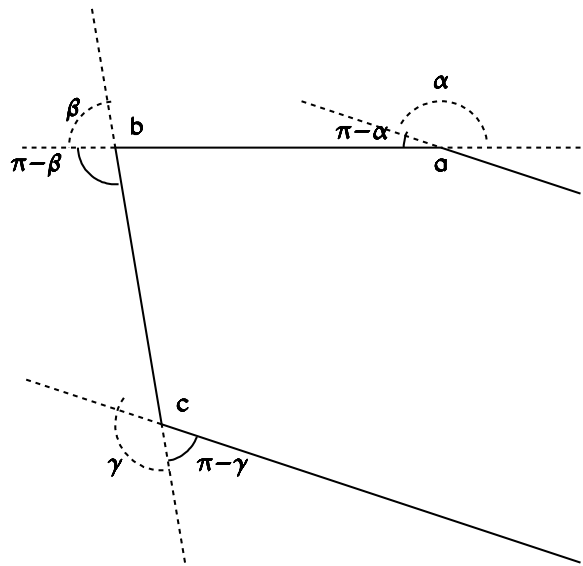}
\caption{Example of Schwartz-Christoffel transformation }
\label{scchr}
\end{figure}
\end{center}

The class of conformal mappings relevant for our application are the Schwartz-Christoffel 
transformations. They map the upper half of the complex $t$-plane onto the interior of a
polygon in such a way that the real axis is mapped onto the boundary.\\
Referring to Fig.\ref{scchr}, the mapping is
\begin{equation}
\frac{dz}{dt} = S(t-a)^{(\alpha/\pi)-1}(t-b)^{(\beta/\pi)-1}(t-c)^{(\gamma/\pi)-1} \cdot
\label{schreq}
\end{equation}
where $S$ is a constant of scale and $a,b,c, \cdots$ are the point of the real axis corresponding
to the polygon vertices.\\
If the polygon is a rectangle, it has four vertices and all angles
equal to $\pi/2$. In this case Eq.\ref{schreq} can be written in integral form as

\begin{equation}
z(t) = S \int_{t_0}^t dt^\prime (t^\prime-a)^{-\frac{1}{2}} (t^\prime-b)^{-\frac{1}{2}}
       (t^\prime-c)^{-\frac{1}{2}} (t^\prime-d)^{-\frac{1}{2}}
\label{schrrect}
\end{equation}

This 4 parameter class of integrals can be reduced to the following one parameter class to be 
discussed in the next section

\begin{equation}
z(t,k) = S^\prime \int_{0}^t dt^\prime \left[ (1-t^{\prime 2}) (1-k^2t^{\prime 2})
       \right] ^{-\frac{1}{2}}
\label{schrell}
\end{equation}


\subsection{Elliptic integrals and functions}

A large class of integrals involving rational 
functions of root square of polynomials up to the fourth degree can be reduced to three types 
of parametrized integrals, the elliptic integrals of first, second and third kinds.\\
The elliptic integral of first kind in its Legendre normal form is the expression 
in Eq.\ref{schrell} with $S^\prime=1$. 
For a general $t$, the integral is said to be {\it incomplete} and is given the symbol $F(t,k)$,
where $k$ is called the {\it modulus}. When $t=1$, it is {\it complete} and is given the symbol
$K(k)=F(1,k)$. Derived from $k$ is the {\it complementary modulus}, $k^\prime = \sqrt{1-k^2}$. 
Another definition is $K^\prime(k) = K(k^\prime)$.\\
In \cite{ellfun} the following important relation is derived

\begin{eqnarray}
F(\pm1/k,k) = \pm K(k)-iK(k^\prime)
\label{ell1ok}
\end{eqnarray}

Other relations deduced from the definition of $F(t,k)$ are

\begin{eqnarray}
F(t,k) - F(\frac{1}{k},k) &=& -F(1,k) + F(\frac{1}{kt},k) \qquad t\ge \frac{1}{k} \nonumber \\
F(t,k) &=& - iK(k^\prime) + F(\frac{1}{kt},k)
\label{ellft}
\end{eqnarray}

Eq.\ref{ellft} reduces the calculation of $F(t,k)$ for $t\ge \frac{1}{k}$ to 
the evaluation of $F(t,k)$ for $t\le 1$. In particular $\lim_{t->\pm \infty} F(t,k) = 
- iK(k^\prime)$.\\
Inverting Eq.\ref{schrell} with $S^\prime=1$, $t$ is expressed as function of $z$ and $k$, 
defining the function sine amplitude $sn$

\begin{equation}
t = sn(z,k) 
\label{ellsin}
\end{equation}

\subsection{Capacitance calculation with Schwartz-Christoffel mapping}

The application of conformal mapping to capacitance calculation in geometry with 2D symmetry 
is based on the fact that capacitance is preserved by conformal transformations. 
For regions with strips 
and planes on the boundary, the mapping is a Schwartz-Christoffel transformation 
\cite{schinziger}-\cite{cap90}-\cite{binns92}.\\
The basic configuration is shown in Fig.\ref{figell}, where the upper half of the $t$-plane 
is mapped into the interior of the rectangle in the $z$-plane.\\
The mapping is expressed by elliptic integrals and functions
\begin{equation}
z = \frac{1}{2K(k)}F(\frac{t}{W_t/2},k)\,\,\,\qquad \qquad k = \frac{W_t}{W_t+2S_t}
\end{equation}
where $W_t$ and $S_t$ are the strip and gap widths in the $t$-plane.\\
This function maps the strip in the $t$-plane onto the upper side of the rectangle in the $z$-plane
and the two lateral half-planes onto the lower side. In the $z$-plane the rectangle width is $W_z=1$ and 
the height is $h_z = \frac{K^\prime(k)}{2K(k)}$.\\
The reverse mapping is 
\begin{equation}
t = \frac{W_t}{2} sn(2K(k) z,k)
\end{equation}
The capacitance between the strip and the planes in the upper half of the $t$-plane is equal
to the parallel plane capacitance in the $z$-plane

\begin{equation}
C = \epsilon   \frac{W_z}{h_z} =  \epsilon \frac{2K(k)}{K^\prime(k)}
\label{capstrpla}
\end{equation}

where $\epsilon$ is the dielectric constant of the material.\\
If the upper and lower half-planes of the $t$-plane are filled with media with different 
dielectric constants $\epsilon_1$ and $\epsilon_2$, the field component normal to the 
non-conducting boundaries is assumed to be zero, so that the calculation is done 
separately for the two half-planes.\\
This approach has been pionereed in \cite{cap90} where several structures approximating a 
micro-strip detector section were studied. The major limitation in the analytical calculation
using conformal transformation consisted in approximating the lateral strips as a single lateral 
half-plane so that $C_n$ could not be estimated separately.\\
This estimation was possible using a mixed analytical-numerical method 
\cite{romania87}-\cite{romania88} that, beyond conformal transformation, involves numerical 
integration and matrix inversion.\\
In this paper this limitation is removed and the relevant capacitances are expressed by 
analytical expressions.

\subsection{Auxiliary structure}

The structure in Fig.\ref{rectangle} is used in the following as an intermediate configuration 
and we need the capacitance of the strip to the opposite plane.\\
This capacitance is calculated applying two transformations. The first one maps the interior 
of the rectangle in the $z$ plane onto the upper half of the $t$-plane

\begin{equation}
t = sn\left(2K(k_1) \frac{z}{W_z+S^l_z+S^r_z},k_1\right)
\end{equation}

where the strip width $W_z$ and the left and right half-gaps $S^l_z$ and $S^r_z$ are defined 
in Fig.{\ref{rectangle}. The modulus $k_1$ is defined by 

\begin{equation}
\frac{W_z+S^l_z+S^r_z}{h_z} = 2\frac{K(k_1)}{K^\prime(k_1)}
\end{equation}

This transcendental equation can be approximately, but with high precision, solved employing a 
series expansion from \cite{ellfun}. The first order is 

\begin{equation}
\frac{K^\prime(k)}{K(k)} = \left\{ \begin{array}{ll}
         \frac{1}{\pi} \ln (2\frac{1+\sqrt{k^\prime}}{1-\sqrt{k^\prime}}) 
& 0\le k \le \frac{1}{\sqrt{2}} \\
         \frac{\pi}{ \ln(2\frac{1+\sqrt{k}}{1-\sqrt{k}}) }
& \frac{1}{\sqrt{2}} \le k \le 1.0
                            \end{array}
\right .
\label{invkoverk}
\end{equation}

that can be inverted to obtain $k$, defining $y = \frac{K^\prime(k)}{K(k)}$, as

\begin{eqnarray}
  k^\prime &=& \left(\frac{\exp^{\pi y}-2.}{\exp^{\pi y}+2.}\right)^2
\qquad\qquad \frac{K^\prime(k)}{K(k)} \ge 1 \nonumber \\
  k &=& \left(\frac{\exp^{\pi/y}-2.}{\exp^{\pi/y}+2.}\right)^2
\qquad\qquad \frac{K^\prime(k)}{K(k)} \le 1 
\label{invk}
\end{eqnarray}

In the $t$-plane the strip width and the half-gaps are

\begin{eqnarray}
W_t &=& sn\left(2K(k_1)\frac{S^l_z+W_z-S^r_z}{W_z+S^l_z+S^r_z},k_1\right) 
      - sn\left(2K(k_1)\frac{S^l_z-W_z-S^r_z}{W_z+S^l_z+S^r_z},k_1\right) \nonumber \\
S^l_t &=& sn\left(2K(k_1)\frac{S^l_z-W_z-S^r_z}{W_z+S^l_z+S^r_z},k_1\right) 
      + \frac{1}{k_1} \nonumber \\
S^r_t &=& \frac{1}{k_1} -
      sn\left(2K(k_1)\frac{S^l_z+W_z-S^r_z}{S^l_zW_z+S^l_z+S^r_z},k_1\right) 
\label{wssag}
\end{eqnarray}

With the definitions 

\begin{eqnarray}
t^l &=& - sn\left(2K(k_1)\frac{S^l_z-W_z-S^r_z}{S^l_zW_z+S^l_z+S^r_z},k_1\right) \nonumber \\
t^r &=& sn\left(2K(k_1)\frac{S^l_z+W_z-S^r_z}{S^l_zW_z+S^l_z+S^r_z},k_1\right)
\label{tlte}
\end{eqnarray}

the mapping from the upper half of the $t$-plane to the interior of the rectangle in the $u$-plane is  

\begin{equation}
u(t) = \int_0^t \frac{dt^\prime}{\sqrt{(t^\prime-t^l) (t^\prime-t^r) (t^\prime-\frac{1}{k_1}) 
       (t^\prime+\frac{1}{k_1})} }
\label{ellu}
\end{equation}

Following the steps outlined in \cite{schinziger}-\cite{ghione83}-\cite{ghione87}, the capacitance is 

\begin{equation}
C = \epsilon \frac{W_u}{h_u} = \epsilon \frac{K^\prime(k_2)}{K(k_2)}\qquad \qquad 
k_2 = \sqrt{\frac{S^l_t}{W_t+S^l_t}} \sqrt{\frac{S^r_t}{W_t+S^r_t}}
\label{capuc}
\end{equation}

\begin{center}
\begin{figure}
\includegraphics[width=1.1\textwidth] {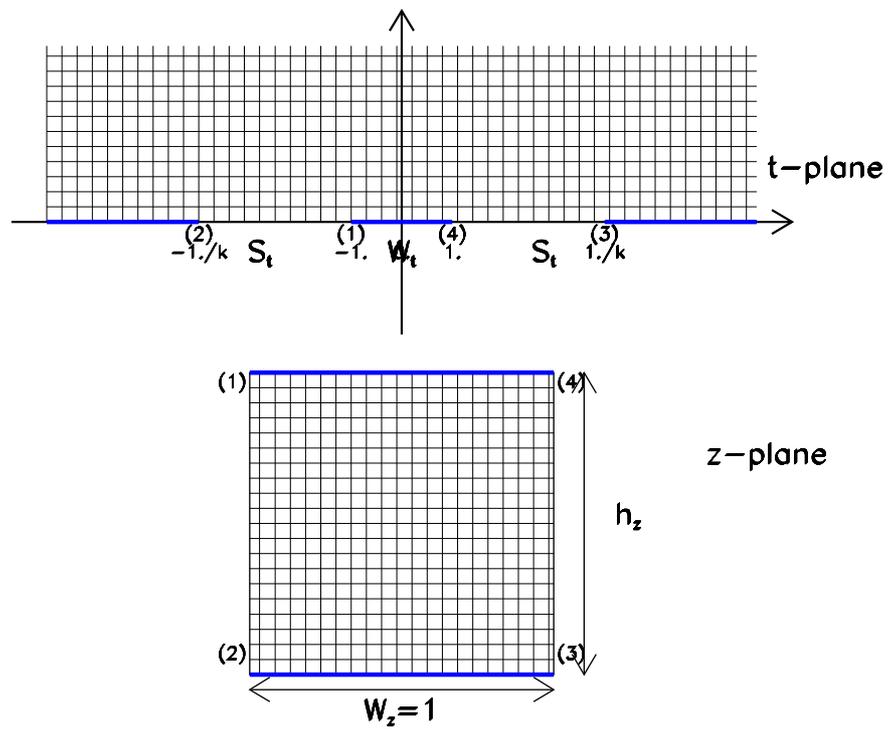}
\caption{Mapping strip with two planes into a rectangular domain. Top: original structure, upper 
half-plane; bottom: parallel plane capacitor}
\label{figell}
\end{figure}
\end{center}

\begin{center}
\begin{figure}
\includegraphics[width=1.1\textwidth] {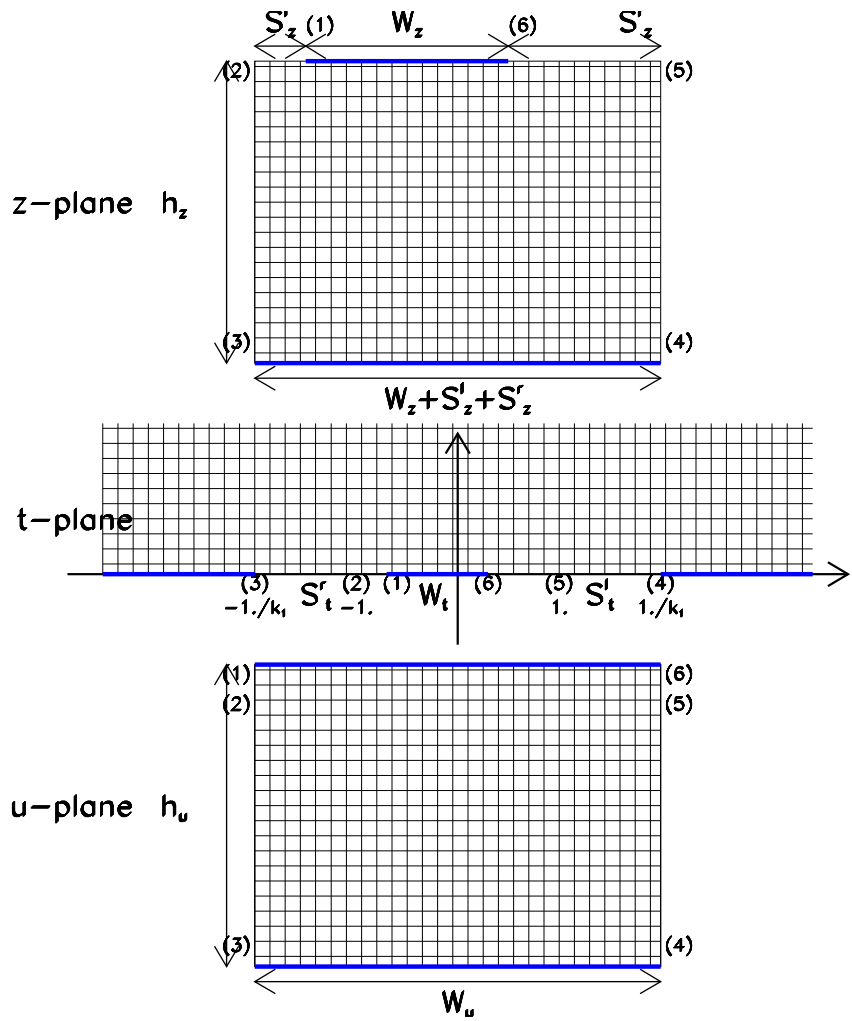}
\caption{Mapping sequence for calculating $C$ in a rectangle with asymmetric gaps. Top: original 
structure; middle: intermediate structure, upper half-plane; bottom: parallel plane capacitor}
\label{rectangle}
\end{figure}
\end{center}

\section{Strip to ground capacitance}

The strip to ground capacitance $C_g$ is calculated applying the 
Schwartz-Christoffel mapping from the rectangular domain in the $z$-plane to the upper
half of the $t$-plane and then the reverse mapping to the rectangular domain in 
the $u$-plane with a different modulus as in Fig.\ref{capgr}.\\
The mapping from $z$ plane to $t$ plane is 

\begin{equation}
t = sn\left(2K(k_1) \frac{z}{W_z+S_z},k_1\right)
\end{equation}

where the modulus $k_1$ is obtained implicitly from

\begin{equation}
\frac{W_z+S_z}{h_z} = 2\frac{K(k_1)}{K^\prime(k_1)}
\label{kone}
\end{equation}
that can be solved using Eq.\ref{invkoverk}-\ref{invk}.

The mapping from $t$ plane to $u$ plane is the elliptic integral

\begin{equation}
u = F\left(\frac{t}{sn\left(K(k_1)\frac{W_z}{W_z+S_z},k_1\right)},k_2\right) 
\label{ellf}
\end{equation}

where $k_2 = k_1 sn(K(k_1)\frac{W_z}{W_z+S_z},k_1)$ and $k_1$ come from Eq.\ref{kone} or 
Eq.\ref{invkoverk}.\\
The capacitance is 

\begin{equation}
C_g = \epsilon \frac{W_u}{h_u} = \epsilon \frac{2K(k_2)}{K^\prime(k_2)}
\label{capug}
\end{equation}

\begin{center}
\begin{figure}
\includegraphics[width=1.1\textwidth] {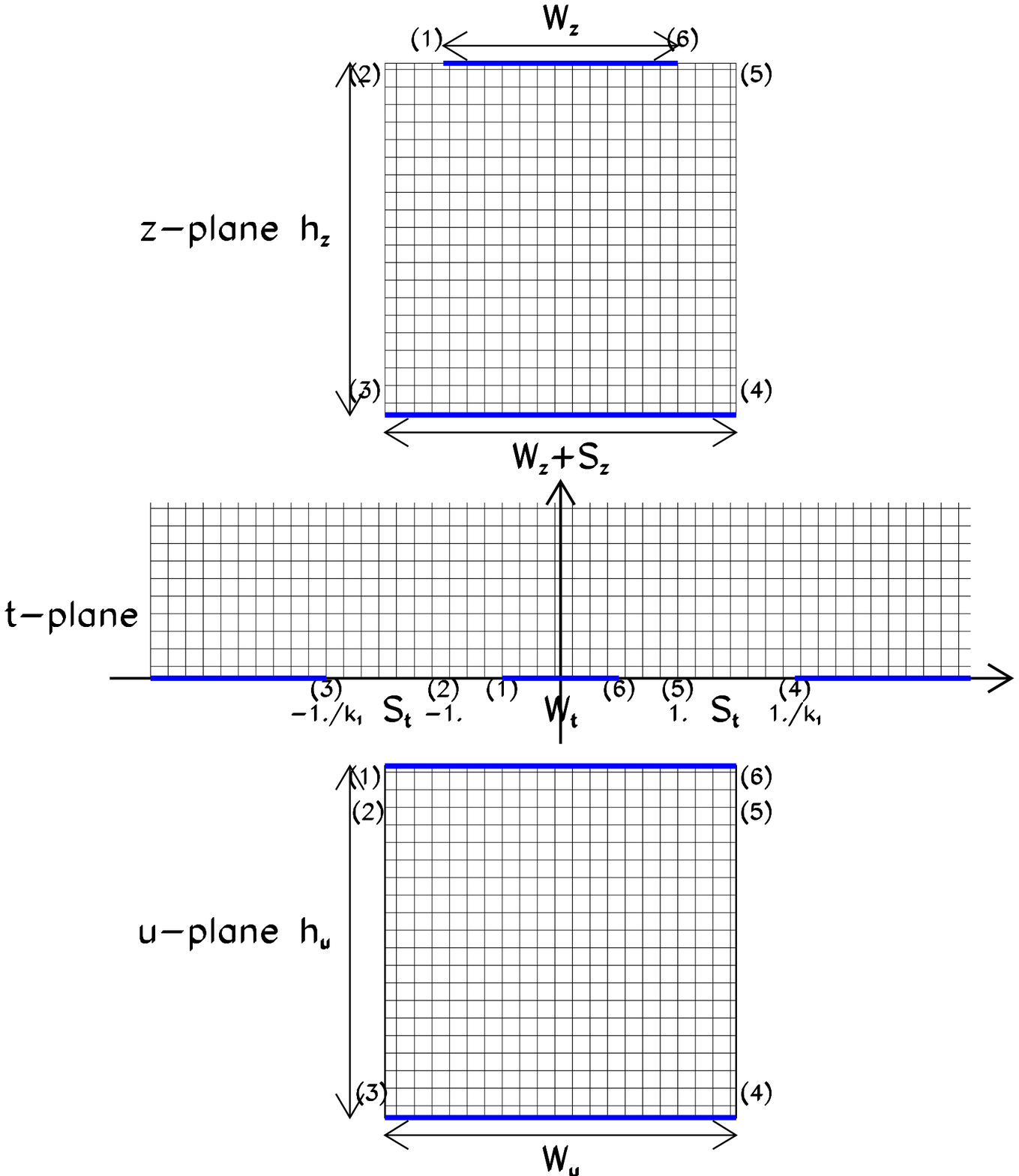}
\caption{Mapping sequence for calculating $C_g$. Top: original structure;
middle: intermediate structure, upper half-plane; bottom: parallel plane capacitor}
\label{capgr}
\end{figure}
\end{center}

In Fig.\ref{capgrplot} $C_g$ normalized to the parallel plate capacitance $C_{pp} = 
\epsilon \frac{W_z+S_z}{h_z}$ is plotted versus the $\frac{W_z}{W_z+S_z}$
with $\frac{W_z+S_z}{h_z}$ as parameter.

\begin{center}
\begin{figure}
\includegraphics[width=1.1\textwidth] {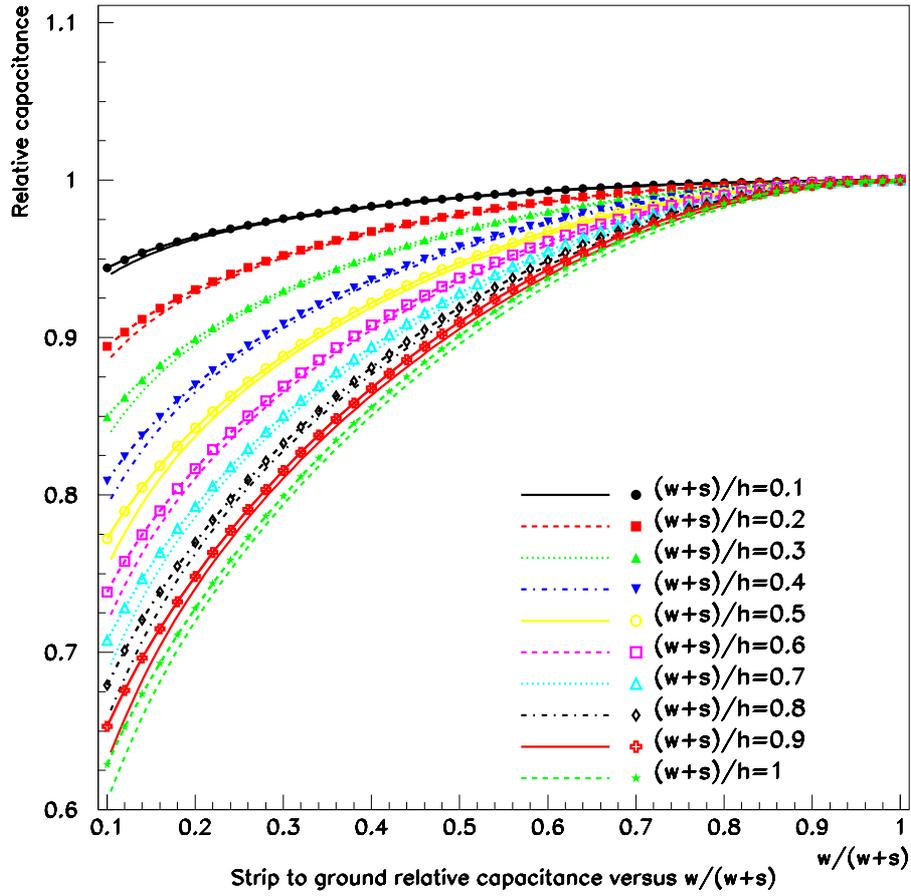}
\caption{$C_g$ normalized to parallel plane capacitance $C_{pp}$ versus $\frac{w}{w+s}$ and $\frac{w+s}{h}$
($w$ is the strip width, $s$ the interstrip gap and $h$ the height).
The curves with symbols are the results of the calculations in this paper.
The curves without symbols are the results of the empirical model proposed in \cite{barb}.}
\label{capgrplot}
\end{figure}
\end{center}

\section{Interstrip capacitances}

\subsection{Preliminary considerations}

The geometry of the detector relevant for the study of interstrip capacitances is displayed in
Fig.\ref{sidecap}. The upper half of the $t$ plane is in air, the lower half in the detector 
material, silicon in the following examples.\\ 
A simplified assumption is that the silicon section extends indefinitely, that is 
$h \rightarrow \infty$. In this case each interstrip capacitance is obtained multiplying the
half-plane capacitance
by $1+\epsilon^{Si}_r$ ($\epsilon^{Si}_r$ is the relative silicon dielectric constant).\\
For later use we assume that the point on each gap where the electric field 
generated by a voltage applied on the central strip is normal to the detector surface is in 
the gap center (interstrip points).
These points must exist for continuity of the electric field and the field lines connecting 
them with the central strip divide the detector in separate regions. 
For each lateral strip, these regions are the volumes where a test positive charge would drift 
under the action of the electric field due to the central strip.\\
In Fig.\ref{sidecap} the coordinates of the strip edges $x_{in}$, $x_{on}$ and the gap centers 
$x_{cn}$ of a small section of the detector are labeled.
The interstrip capacitances in the upper half-plane are estimated with the help of 
Fig.\ref{sideaircap}. $x_{cn}$ are the interstrip points.\\
The points marked on the picture are:
\begin{eqnarray}
x_{in} & = & n(W_t+S_t) - W_t/2\\
x_{on} & = & n(W_t+S_t) + W_t/2\\
x_{cn} & = &(n+1/2)(W_t+S_t)  
\end{eqnarray}

The reason for introducing these interstrip points is that it can be proved that in the rectangular 
domain the field lines running from these points to the opposite plane are straight lines normal 
to the planes. 
Those divide the rectangular domain in smaller rectangles with the geometry shown in Fig.\ref{rectangle}
in which the Poisson equation can be solved separately. 

\subsection{Capacitances between neighbouring and not-neighbouring strips}

In Fig.\ref{sideaircap} the upper half of the $t$-plane is mapped onto the interior of the rectangle
in the $u$-plane by the transformation 
\begin{equation}
u = \frac{1}{K(k_1)}F\left(\frac{t}{W_t/2},k_1\right) \qquad \qquad k_1 = \frac{W_t}{W_t+2s_t}
\end{equation}
The central strip in the $t$ plane is mapped onto the upper side in the $u$ plane with width
$W_u=2$. The gaps on the side of the central strip are mapped onto the lateral sides of 
the rectangle.\\
Eq.\ref{ell1ok} gives

\begin{equation} 
h_u=K(k_1^\prime)
\label{hu}
\end{equation} 

and Eq.\ref{ellft} implies that all lateral strips are mapped onto strips on the lower side 
of the rectangle.\\
The $n-th$ lateral strip is mapped onto a strip with width
\begin{eqnarray}
W_{un} &=& F(x_{on}/(W_t/2),k_1)-F(x_{in}/(W_t/2),k_1) \nonumber \\
&=& F(-\frac{1}{k_1}+(2n+1)\frac{W_t+S_t}{W_t},k_1) - F(\frac{1}{k_1}+(2n-1)\frac{W_t+S_t}{W_t},k_1))
\label{wun}
\end{eqnarray}
The left and right gaps of the $n-th$ lateral strip are mapped onto gaps with widths 
(defining $n=1$ $F(x_{cn-1}/(W_t/2),k_1))=0$)

\begin{eqnarray}
S^l_{un} &=& F(x_{cn}/(W_t/2),k_1)-F(x_{on}/(W_t/2),k_1) \nonumber \\ 
&=& F((2n+1)\frac{W_t+S_t}{W_t},k_1) - F(-\frac{1}{k}+(2n+1)\frac{W_t+S_t}{W_t},k_1)
\label{slun}
\end{eqnarray}

\begin{eqnarray}
S^r_{un} &=& F(x_{in}/(W_t/2),k_1)- F(x_{cn-1}/(W_t/2),k_1) \nonumber \\
&=& F(\frac{1}{k}+(2n-1)(\frac{W_t+S_t}{W_t}),k_1) - F((2n+1)\frac{W_t+S_t}{W_t},k_1) 
\label{srun}
\end{eqnarray}

The problem of calculating $C_n$ in a half-plane is reduced to the calculation of the strip 
to backplane capacitance in a structure like Fig.\ref{rectangle}, following the steps described 
previously. The contributions in air and in silicon are added.\\
The steps are the followung: a modulus $k_1$ is determined by 

\begin{equation}
\frac{W_{un}+S^l_{un}+S^r_{un}}{h_u} = 2\frac{K(k_1)}{K^\prime(k_1)}
\end{equation}

using Eq.\ref{invkoverk}. Eq.\ref{wssag} gives

\begin{eqnarray}
W_n &=& sn\left(2K(k_1)\frac{S^l_{un}+W_{un}-S^r_{un}}{W_{un}+S^l_{un}+S^r_{un}},k_1\right) 
      - sn\left(2K(k_1)\frac{S^l_{un}-W_{un}-S^r_{un}}{W_{un}+S^l_{un}+S^r_{un}},k_1\right) \\
S^l_n &=& sn\left(2K(k_1)\frac{S^l_{un}-W_{un}-S^r_{un}}{W_{un}+S^l_{un}+S^r_{un}},k_1\right) 
      + \frac{1}{k_1} \\
S^r_n &=& \frac{1}{k_1} -
      sn\left(2K(k_1)\frac{S^l_{un}+W_{un}-S^r_{un}}{S^l_{un}W_{un}+S^l_{un}+S^r_{un}},k_1\right) \\
\label{wssagn}
\end{eqnarray}
and therefore the $n-th$ order interstrip capacitance is 

\begin{equation}
C_n = \epsilon \frac{2K(k^n_2)}{K^\prime(k^n_2)}\qquad \qquad 
k^n_2 = \sqrt{\frac{S^l_{n}}{W_{n}+S^l_{n}}} \sqrt{\frac{S^r_{n}}{W_{n}+S^r_{n}}}
\label{capun}
\end{equation}

The interstrip capacitance $C_1$ between neighboring strips is displayed in Fig.\ref{plotcis1}
versus $W/(W+S)$. The higher order interstrip capacitances $C_n$ 
for $n \in (2,7)$ are displayed in Fig.\ref{plotcis27}. $C_1$ depends strongly on $W/(W+S)$ while 
the $C_n$ dependence is much weaker.\\
Fig.\ref{plotcistot} displays the total interstrip capacitance $C_{is,tot}$ obtained adding all 
interstrip capacitances up to the 7-th order on both sides and the total capacitance 
$C_{tot}=C_{is,tot}+C_g$.
These quantities are easier to measure than $C_n$ and to compare 
with experimental data.

\begin{center}
\begin{figure}
\includegraphics[width=1.1\textwidth] {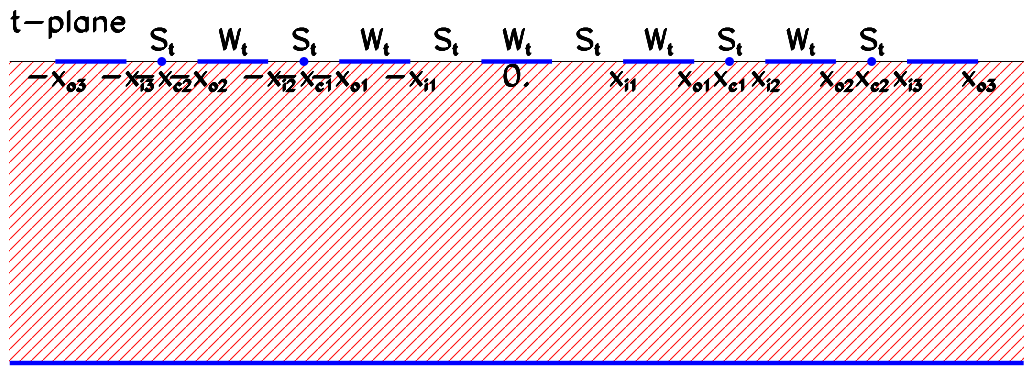}
\caption{Section of a micro-strip detector with strip edges and interstrip gap centers marked.}
\label{sidecap}
\end{figure}
\end{center}

\begin{center}
\begin{figure}
\includegraphics[width=1.1\textwidth] {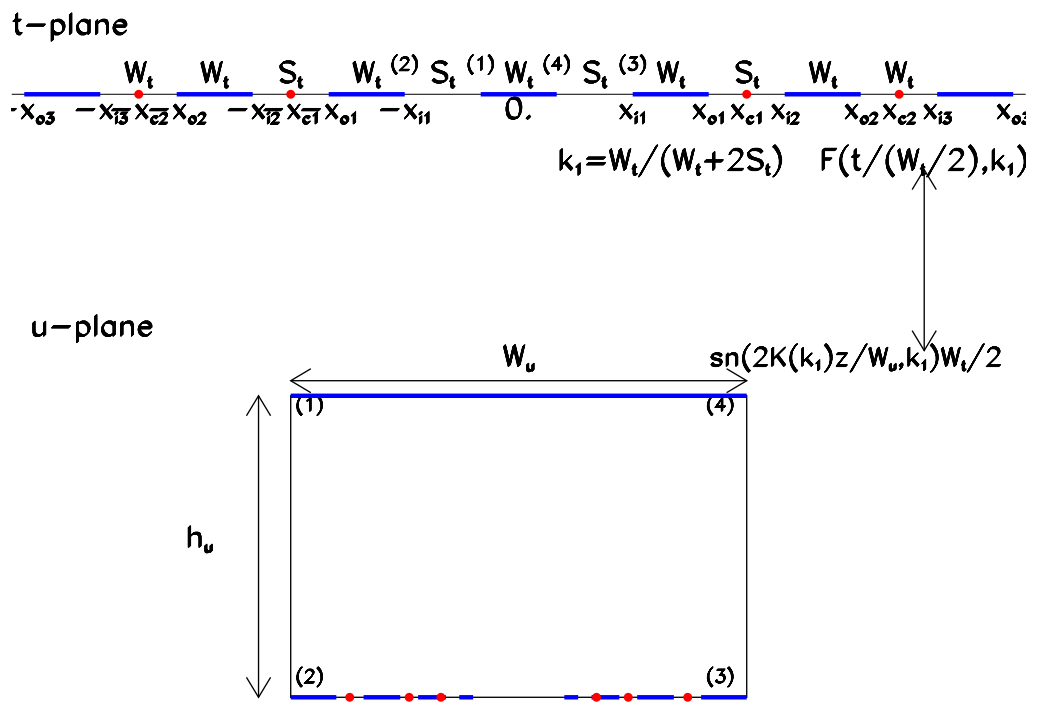}
\caption{Mapping of the half-plane in air. Top: original structure with few strips, the strips 
extends indefinitely on both sides. Bottom: rectangular domain with all lateral strips mapped 
on the lower side.}
\label{sideaircap}
\end{figure}
\end{center}

\begin{center}
\begin{figure}
\includegraphics[width=1.3\textwidth] {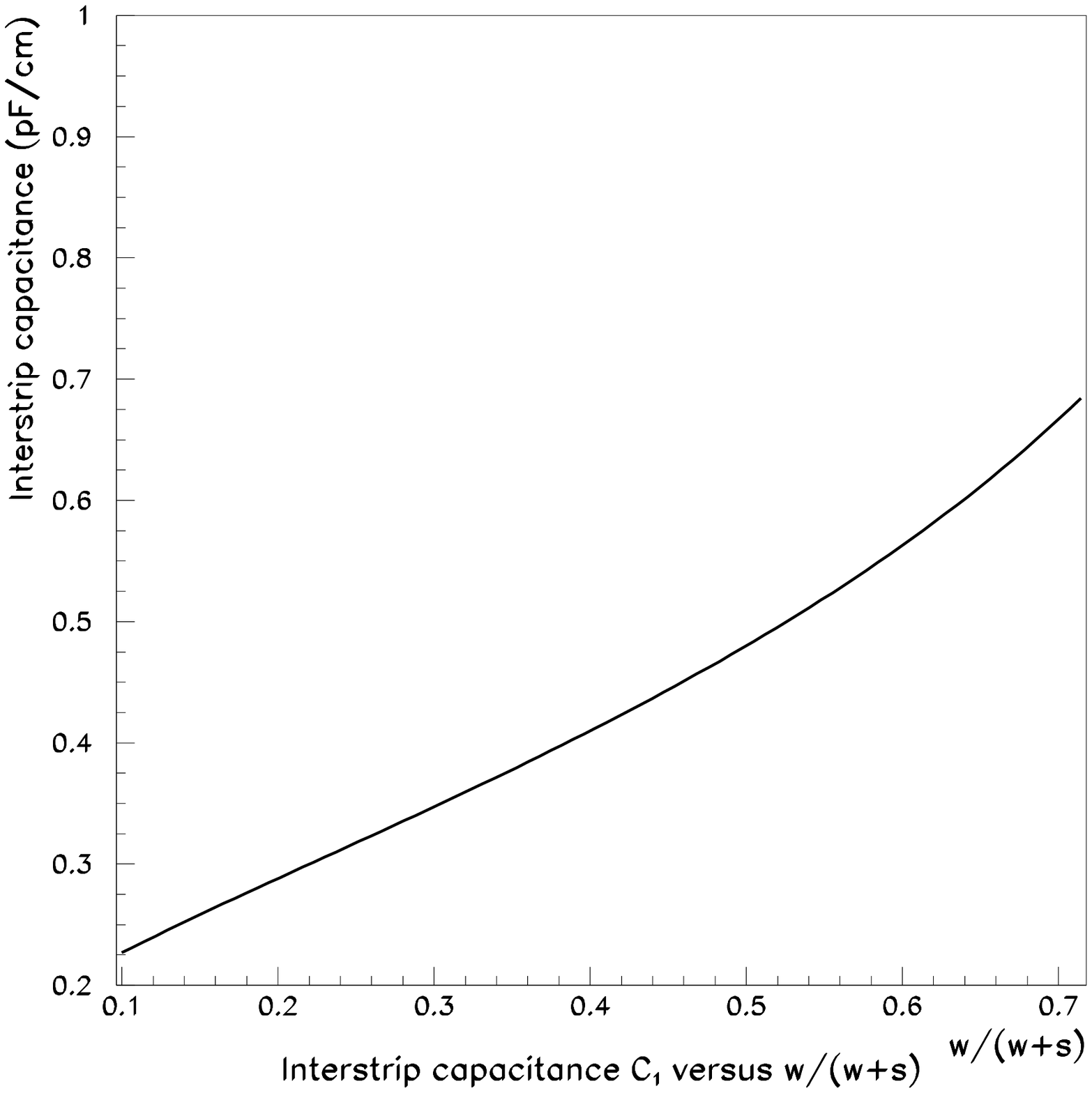}
\caption{Plot of the first order interstrip capacitance $C_1$ versus $w/(w+s)$
($w$ is the strip width, $s$ the interstrip gap and $h$ the height).
}
\label{plotcis1}
\end{figure}
\end{center}

\begin{center}
\begin{figure}
\includegraphics[width=1.3\textwidth] {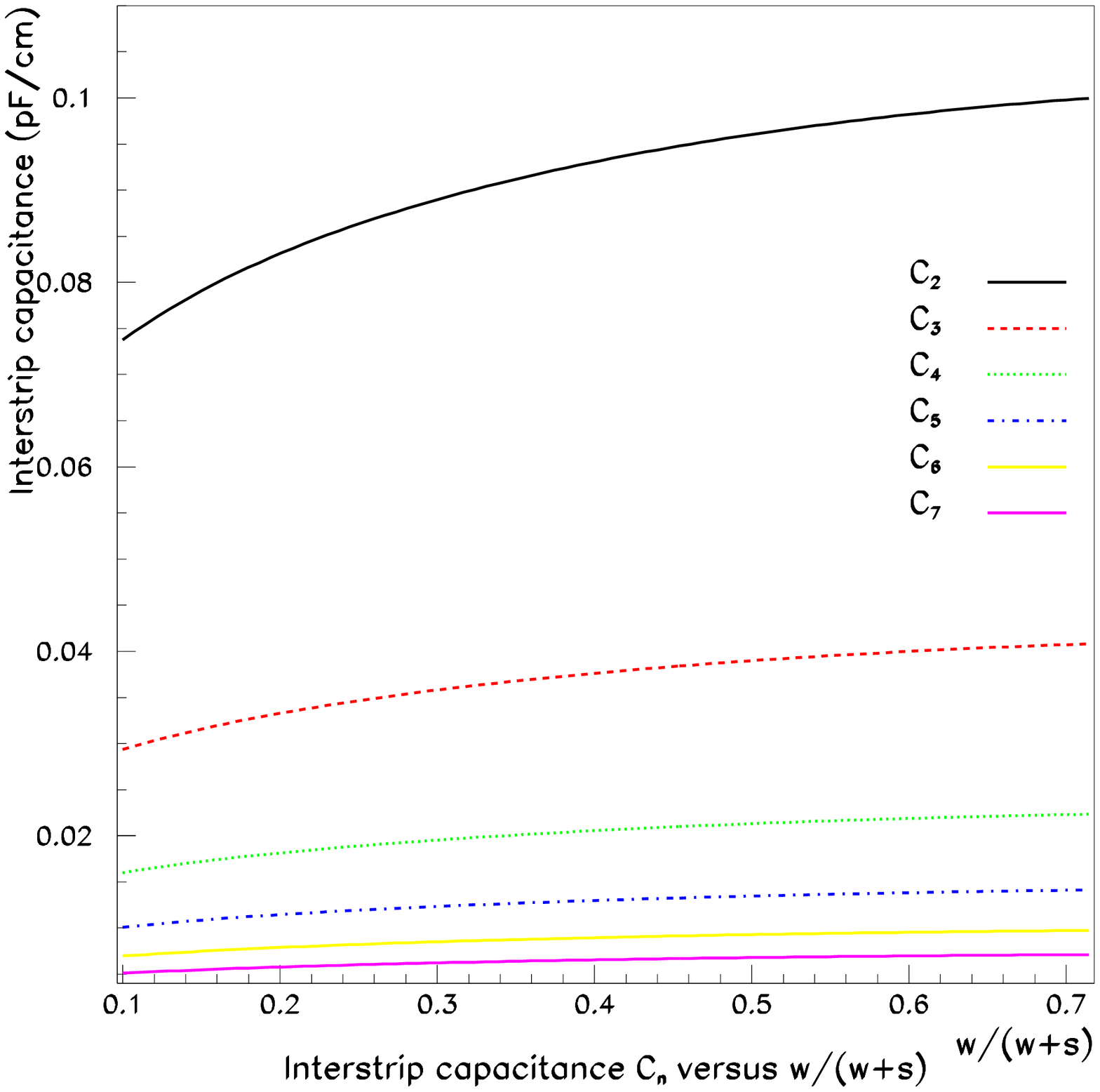}
\caption{Plot of the higher order interstrip capacitances $C_n$ versus $w/(w+s)$
($w$ is the strip width, $s$ the interstrip gap and $h$ the height).
}
\label{plotcis27}
\end{figure}
\end{center}

\begin{center}
\begin{figure}
\includegraphics[width=1.3\textwidth] {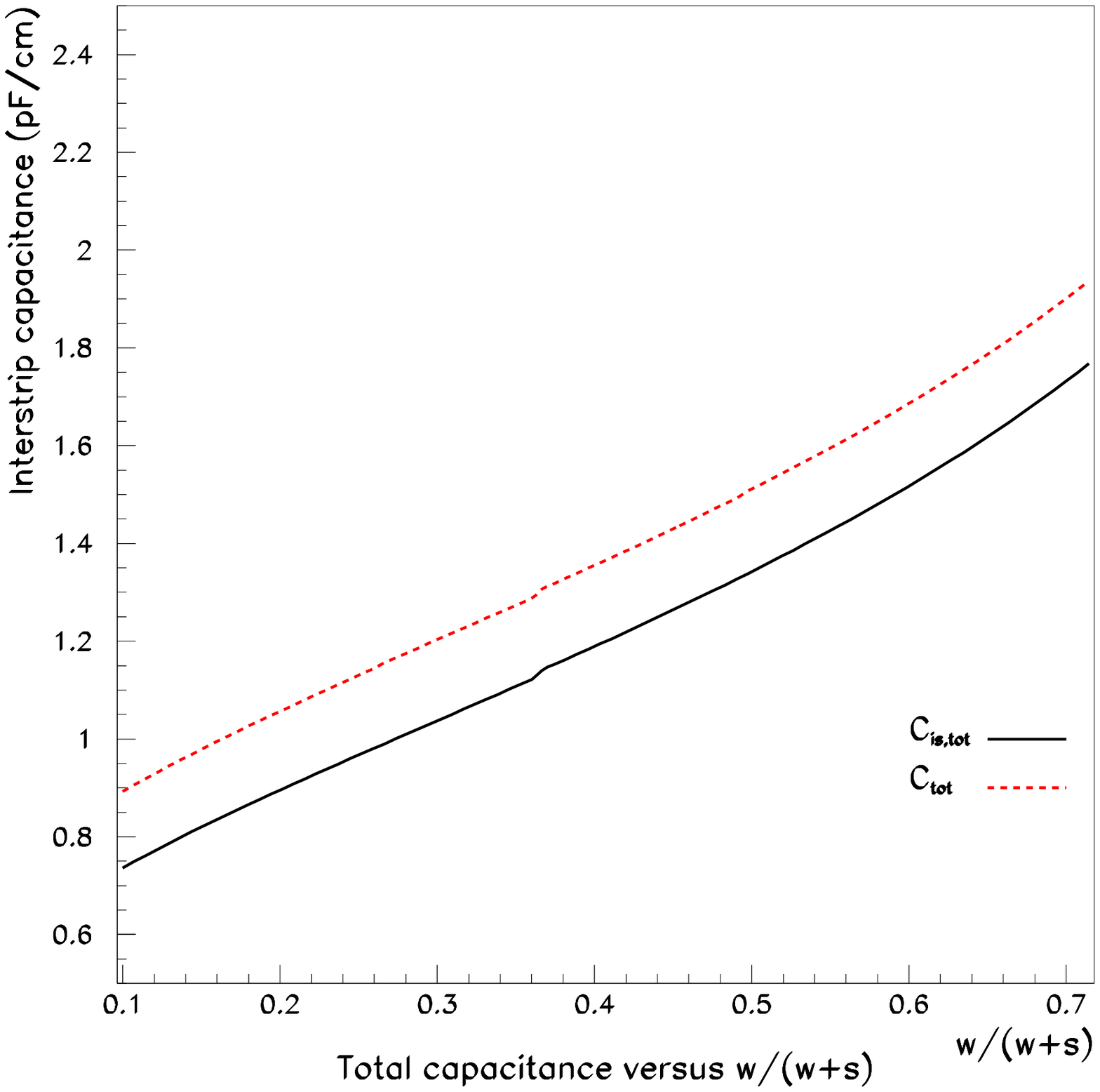}
\caption{Plot of the total interstrip $C_{is,tot}$ and total $C_{tot}$ 
capacitances versus $w/(w+s)$ for $w+s=50\mu m$ and $h=300\mu m$.
($w$ is the strip width, $s$ the interstrip gap and $h$ the height).
}
\label{plotcistot}
\end{figure}
\end{center}

\section{Comparison with other approaches and experimental results}

The results presented so far refer to geometrical capacitances in fully depleted detectors
with strips on the junction side.
Realistic detectors may differ from this model for various reasons. The most relevant is the 
presence of free charges below the interstrip oxide acting as additional conducting layers.
This effect is difficult to predict and describe and may alter significantly the measurements.\\
For detectors not satisfying the previous conditions, either our model is made more realistic, 
and more difficult to solve analytically, or a numerical approach like that in \cite{stripcalc}
can complement our analytical approach by studying the effects of deviation from ideality.\\
In the following, the field of applicability of the calculations developed in 
this paper is studied by comparing them with other models and the experimental results presented
in \cite{barb,cmsstrip}. These results have been selected because they refer to fully depleted 
detectors with strips on the junction side and scan a broad range of detector parameters. 
Furthermore data are presented in a format adequate for comparison with calculations.

\subsection{$C_g$ comparison}

In Fig.\ref{capgrplot} the values of $C_g$ calculated with our model are compared with the empirical 
formula developed in \cite{barb} obtained fitting the results of the numerical solution of the Poisson
equation presented in \cite{santacruz}. This formula is expressed as
\begin{equation}
\frac{C_g}{C_{pp}} = \frac{1}{1+\frac{W+S}{h}f(\frac{W}{W+S})}
\label{empfor}
\end{equation}
where $f(\frac{W}{W+S})$ is defined in \cite{barb}. This model is supported by experimental results 
presented in \cite{barb,cmsstrip}.\\
In \cite{cmsstrip} results for different material orientations $\langle 111\rangle $ and $\langle 100\rangle $ are presented.
$\langle 111\rangle $ has an higher trapped charge at the Si-O$_2$ interface and is more sensitive 
to irradiation.
Nevertheless for nonirradiated devices, in the frequency range relevant for detector signal and
for adequately high overdepletion voltage the capacitances are not dependent on the material orientation
but only on the detector geometry and can be compared to our calculations.\\
Fig.\ref{capgrplot} shows an excellent agreement between our results and Eq.\ref{empfor}.
Therefore, at least for $C_g$, detectors can be operated in such a condition that
the geometrical capacitances are experimentally measurable and they are described very well by our 
calculations.

\subsection{$C_n$ comparison}

A comparison of $C_n$ is possible with the model developed in \cite{cap90}, where the $C_n$ for
$n \in (2,7)$ were calculated for $W=S$. The results are compared in Tab.\ref{cncomp} showing
an excellent agreement

\begin{table}[htb]
\begin{center}
\caption{$C_n$(fF/cm) calculated in \cite{cap90} and in this paper for $W=S$}
\label{cncomp}
\begin{tabular}{|l|l|l|l|l|l|l|l|}
\hline
& $C_1$ & $C_2$ & $C_3$ & $C_4$ & $C_5$ & $C_6$ & $C_7$ \\
\hline
\cite{cap90}  & 487. & 96. & 40. & 22. & 13. & 9. & 6. \\
\hline
Eq.\ref{capun} & 478. & 95. & 39. & 21. & 13. & 9. & 7.\\
\hline
\end{tabular}
\end{center}
\end{table}

Measuring separately $C_n$ is quite difficult. The most straightforward comparison is with the 
total capacitance $C_{tot}$ (including $C_g$) or with the total interstrip capacitance $C_{is,tot}$.

\subsection{$C_{tot}$ comparison}

Experimental measurements of $C_{tot}$ are presented in \cite{barb,cmsstrip}. In order to compare the
results, it is necessary to fix also $(W+S)/h$. $C_{tot}$ and $C_{is,tot}$ shown in Fig.\ref{plotcistot}
are calculated with $W+S=50\mu m$ and $h=300\mu m$. The easiest comparison is with the formula, supported 
by measurements on fully depleted detectors with strips on the junction side, presented in \cite{cmsstrip}, 
\begin{equation}
C_{tot} = 0.8 + 1.6 \frac{w}{w+s} pF/cm
\label{ctotfor}
\end{equation}

A linear fit to $C_{tot}$ in Fig.\ref{plotcistot} (not shown) gives

\begin{equation}
C_{tot} = 0.73 + 1.60 \frac{w}{w+s} pF/cm
\label{ctotfit}
\end{equation}

showing an excellent agreement.

\section{Conclusions}

The capacitances in micro-strip detectors are calculated analytically with the help of
Schwartz-Christoffel conformal transformations. The results are expressed with elliptic functions.\\
The strip to ground, the total and the interstrip capacitances are calculated 
as function of the geometry of the detector.\\
The results are compared with other models and with empirical formulae supported by experimental 
measurements showing an excellent agreement for nonirradiated and fully depleted detectors.

\section*{Appendix}

The numerical evaluation of the complete and incomplete elliptic integrals and of the elliptic functions
is based on FORTRAN 77 routines from the CERN library package available in \cite{cernlib}. 
These routines are based on numerical algorithms developed in \cite{ellnum1, ellnum2, ellnum3, ellnum4}.

\bibliographystyle{unsrt}
\bibliography{ssecap}

\begin{thebibliography}{10}

\bibitem{cap90}
P.~W. Cattaneo.
\newblock Capacitance calculation in a microstrip detector and its application
  to signal processing.
\newblock {\em Nucl. Instr. and Meth. A}, 295:207, 1990.

\bibitem{cat95}
P.~W. Cattaneo.
\newblock Noise and signal processing in a microstrip detector with a time
  variant readout system.
\newblock {\em Nucl. Instr. and Meth. A}, 359:551, 1995.

\bibitem{binns}
K.~J. Binns and P.~J. Lawrenson.
\newblock {\em Analysis and Computation of Electric and Magnetic Field
  Problems}.
\newblock Pergamon, Oxford, 1963.

\bibitem{schinziger}
R.~Schinziger and P.~A.~A. Laura.
\newblock {\em Conformal Mapping: Methods and Applications}.
\newblock Elsevier, Amsterdam, 1991.

\bibitem{alfhors}
L.~Alfhors.
\newblock {\em Complex analysis}.
\newblock McGraw-Hill, Singapore, 1979.

\bibitem{ellfun}
F.~Oberhettinger and W.~Magnus.
\newblock {\em Anwendung der Elliptische Funktionen in Physik und Technik}.
\newblock Springer, Berlin, 1949.

\bibitem{binns92}
K.~J. Binns, P.~J. Lawrenson, and C.~W. Trowbridge.
\newblock {\em The analytical and numerical solution of electric and magnetic
  fields}.
\newblock Wiley, Chichester, England, 1992.

\bibitem{romania87}
D.~Homentcovschi, A.~Manolescu, A.~Manuela, and C.~Burileanu.
\newblock A general approach to analysis of distributed resistive structures.
\newblock {\em IEEE Transactions on Electron Device}, ED-25:787, 1987.

\bibitem{romania88}
D.~Homentcovschi, A.~Manolescu, A.~Manuela, and L.~Kreindler.
\newblock An analytical solution for the coupled stripline-like microstrip line
  problem.
\newblock {\em IEEE Transactions on Microwave Theory and Technology},
  MTT-36:1003, 1988.

\bibitem{ghione83}
G.~Ghione and C.~Naldi.
\newblock Parameters of coplanar waveguides with lower ground plane, 1983.

\bibitem{ghione87}
G.~Ghione and C.~Naldi.
\newblock Coplanar waveguides for {MMIC} applications: Effect of upper
  shielding, conductor backing, finite-extent ground planes, and line-to-line
  coupling.
\newblock {\em IEEE Transactions on Microwave Theory and Technology},
  MTT-29:260, 1987.

\bibitem{barb}
E.~Barberis {et al.}
\newblock Capacitances in silicon microstrip detectors.
\newblock {\em Nucl. Instr. and Meth. A}, 342:90, 1994.

\bibitem{stripcalc}
S.~Chatterji et~al.
\newblock Analysis of interstrip capacitance of si microstrip detector using
  simulation approach.
\newblock {\em Solid State Electronics}, 47:1491, 2003.

\bibitem{cmsstrip}
S.~Albergo et~al.
\newblock Comparative study of $<$111$>$ and $<$100$>$ crystals and capacitance
  measurements on {S}i strip detectors in {CMS}.
\newblock {\em Il Nuovo Cimento}, 112A(11):1261, 1999.

\bibitem{santacruz}
R.~Sonnenblick et~al.
\newblock Electrostatic simulations for the design of silicon strip detectors
  and front-end electronics.
\newblock {\em Nucl. Instr. and Meth. A}, 310:189, 1991.

\bibitem{cernlib}
{\em CERNLIB Short Writeups}, 1996.

\bibitem{ellnum1}
R.~Bulirsch.
\newblock Numerical calculations of elliptic integrals and elliptic functions.
\newblock {\em Numerische Mathematik}, 7:78--90, 1965.

\bibitem{ellnum2}
R.~Bulirsch.
\newblock Numerical calculations of elliptic integrals and elliptic functions
  {II}.
\newblock {\em Numerische Mathematik}, 7:353--354, 1965.

\bibitem{ellnum3}
R.~Bulirsch.
\newblock Numerical calculations of elliptic integrals and elliptic functions
  {III}.
\newblock {\em Numerische Mathematik}, 13:305--315, 1969.

\bibitem{ellnum4}
W.~J. Cody.
\newblock Chebyshev approximations for the complete elliptic integrals {K} and
  {E}.
\newblock {\em Mathematics of Computation}, 19:105--112, 1965.

\end{thebibliography}

\end{document}